\documentclass[5p,times,twocolumn]{elsarticle}

\usepackage[T1]{fontenc}
\usepackage[utf8]{inputenc}
\usepackage{amsmath,amssymb,bm}
\usepackage{graphicx}
\usepackage{booktabs}
\usepackage{array}
\usepackage{xspace}
\usepackage[colorlinks=true,linkcolor=blue,citecolor=blue,urlcolor=blue]{hyperref}

\journal{Physics Letters B}
\biboptions{sort&compress}

\newcommand{\GeV}{\,\mathrm{GeV}}

\newcommand{\msbar}{\overline{\mathrm{MS}}}

\newcommand{\rhophys}{\rho_{Zt}}
\newcommand{\rhowt}{\rho_{Wt}}
\newcommand{\rhomat}{\widehat\rho_{Zt}}

\begin{document}

\begin{frontmatter}

\title{\Large
The Higgs--top--$Z$ mass coincidence relation after NNLO matching}

\author[um]{E. Torrente-Lujan}
\ead{etl@um.es}
\address[um]{Departamento de Fisica, Campus de Espinardo, Universidad de Murcia, \\
30100 Murcia, Spain}

\begin{abstract}
The relation $M_H^2\simeq M_ZM_t$, previously proposed as a Higgs mass coincidence, is reconsidered with present electroweak inputs and with a scheme-consistent matching analysis.  With the 2025 PDG values for $M_Z$, $M_W$ and $M_H$, and the ATLAS--CMS direct top-mass combination, the pole-level ratio is
$\rho_{Zt}=M_ZM_t/M_H^2=1.00362\pm0.00261$.
Thus an exact pole-level geometric relation predicts either $M_H=125.426\pm0.120\,\mathrm{GeV}$ or $M_t=171.898\pm0.302\,\mathrm{GeV}$, which is still a $1.4\sigma$ test rather than an exclusion.  By contrast, the companion arithmetic relation gives $\rho_{Wt}=(M_W+M_t)/(2M_H)=1.00994\pm0.00159$ and is not a viable exact mass sum rule.  We then evaluate the complete NNLO weak-scale $\overline{\mathrm{MS}}$ matching formulae at $\mu=M_t$.  In the standard convention one obtains
$\widehat\rho_{Zt}(M_t)=\sqrt{g_2^2+g_Y^2}\,y_t/(4\sqrt2\lambda)=0.96714\pm0.00361$.
Consequently, the exact running-coupling boundary condition $\lambda=g_Zy_t/(4\sqrt2)$ at the top scale would predict $M_H=123.19\pm0.20\,\mathrm{GeV}$, or equivalently $M_t=177.81\pm0.50\,\mathrm{GeV}$ when $M_H$ is held fixed.  This is incompatible with the measured point.  A possible symmetry explanation must therefore act on pole-level threshold quantities, or provide a finite matching factor $\kappa_{\rm th}=1.0340\pm0.0039$ at the electroweak scale.  We formulate this requirement as a target for custodial/top-Higgs or triality-like symmetry extensions.
\end{abstract}


\end{frontmatter}

\section{Motivation}
\label{sec:motivation}

After the Higgs discovery the numerical observation
\begin{equation}
  M_H^2\simeq M_ZM_t
  \label{eq:pole_relation}
\end{equation}
was proposed as a possible electroweak mass coincidence involving the heaviest spin-0, spin-1/2 and spin-1 representatives of the Standard Model (SM) spectrum \cite{Torrente2014}.  A second relation,
\begin{equation}
  2M_H\simeq M_W+M_t,
  \label{eq:arithmetic_relation}
\end{equation}
was also close at early LHC precision.  Present data allow one to decide which of the old numerical observations can still be used as a boundary-condition target, and which are already excluded as exact relations.

We use capital letters for physical pole-like masses.  The direct LHC top mass is treated as a pole-mass proxy when evaluating Eqs.~(\ref{eq:pole_relation}) and (\ref{eq:arithmetic_relation}); the corresponding theoretical caveat is discussed below.  Running $\overline{\rm MS}$ couplings at a scale $\mu$ are denoted by $g_2(\mu)$, $g_Y(\mu)$, $y_t(\mu)$ and $\lambda(\mu)$, with $g_Z=(g_2^2+g_Y^2)^{1/2}$.

\section{Updated pole-level tests and predictions}
\label{sec:pole}

The numerical inputs are collected in Table~\ref{tab:inputs}.  For the strong coupling used in the matching section we take the PDG 2025 value $\alpha_s(M_Z)=0.1180\pm0.0009$.

\begin{table}[t]
\centering
\caption{Mass inputs.  The quoted $M_W$ value is the PDG 2025 world average excluding the CDF Run-II result; replacing it by the 2026 CMS value changes only the last digits of $\rho_{Wt}$.}
\label{tab:inputs}
\begin{tabular}{ll}
\toprule
Quantity & Value \\
\midrule
$M_Z$ & $91.1880\pm0.0020\GeV$ \\
$M_W$ & $80.3692\pm0.0133\GeV$ \\
$M_H$ & $125.20\pm0.11\GeV$ \\
$M_t$ & $172.52\pm0.33\GeV$ \\
$\alpha_s(M_Z)$ & $0.1180\pm0.0009$ \\
\bottomrule
\end{tabular}
\end{table}

The two dimensionless ratios are
\begin{align}
\rhophys&\equiv\frac{M_ZM_t}{M_H^2}=1.00362\pm0.00261,\label{eq:rho_pole}\\
\rhowt&\equiv\frac{M_W+M_t}{2M_H}=1.00994\pm0.00159.\label{eq:rho_arith}
\end{align}
The corresponding logarithmic residuals are
\begin{align}
\Delta_{Zt}&\equiv \ln\rhophys=(3.61\pm2.60)\times10^{-3},\label{eq:delta_pole}\\
\Delta_{Wt}&\equiv \ln\rhowt=(9.89\pm1.57)\times10^{-3}.\label{eq:delta_arith}
\end{align}
Thus Eq.~(\ref{eq:pole_relation}) remains compatible with exact unity at about $1.4\sigma$, while Eq.~(\ref{eq:arithmetic_relation}) is about $6.3\sigma$ away from exact unity.  The recent CMS measurement $M_W=80.3602\pm0.0099\GeV$ gives $\rho_{Wt}=1.00990\pm0.00159$, so the conclusion about the arithmetic relation is unchanged.

Table~\ref{tab:predictions} shows the falsifiable predictions obtained when a relation is imposed as an exact equation.  The geometric relation predicts a top mass slightly below the present direct average.  This is the most useful pole-level prediction, because the future limiting uncertainty is expected to be the top-mass definition rather than $M_Z$ or $M_H$.  The arithmetic relation predicts a top mass near $170.03\GeV$ and is already excluded as an exact mass sum rule.

\begin{table}[t]
\centering
\caption{Predictions obtained by imposing exact relations.  The last two rows use the $\overline{\rm MS}$ boundary condition discussed in Sec.~\ref{sec:matching}.}
\label{tab:predictions}
\begin{tabular}{lll}
\toprule
Exact condition & Predicted quantity & Result \\
\midrule
$M_H^2=M_ZM_t$ & $M_H$ & $125.426\pm0.120\GeV$ \\
$M_H^2=M_ZM_t$ & $M_t$ & $171.898\pm0.302\GeV$ \\
$2M_H=M_W+M_t$ & $M_H$ & $126.445\pm0.165\GeV$ \\
$2M_H=M_W+M_t$ & $M_t$ & $170.031\pm0.220\GeV$ \\
$\lambda=g_Zy_t/(4\sqrt2)$ & $M_H$ & $123.19\pm0.20\GeV$ \\
$\lambda=g_Zy_t/(4\sqrt2)$ & $M_t$ & $177.81\pm0.50\GeV$ \\
\bottomrule
\end{tabular}
\end{table}

\begin{figure}[t]
\centering
\includegraphics[width=\columnwidth]{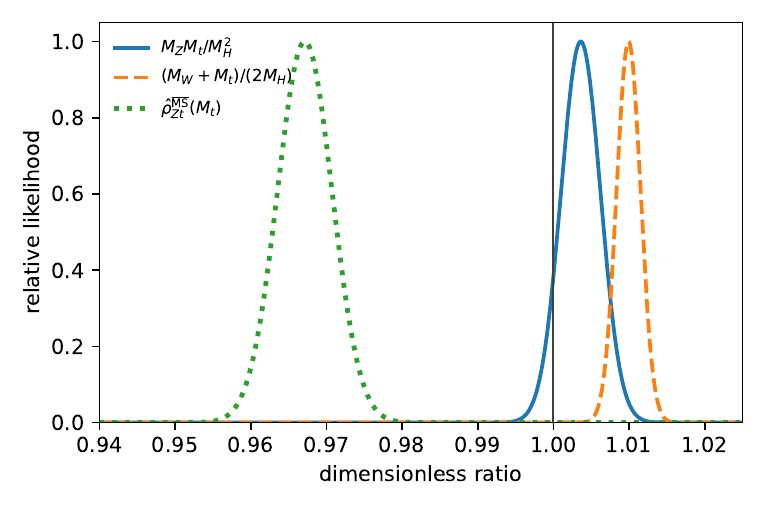}
\caption{Relative Gaussian likelihoods for the pole-level geometric ratio, the pole-level arithmetic ratio, and the running-coupling ratio after NNLO $\overline{\rm MS}$ matching at $\mu=M_t$.  The vertical line is exact unity.}
\label{fig:likelihoods}
\end{figure}

\begin{figure}[t]
\centering
\includegraphics[width=\columnwidth]{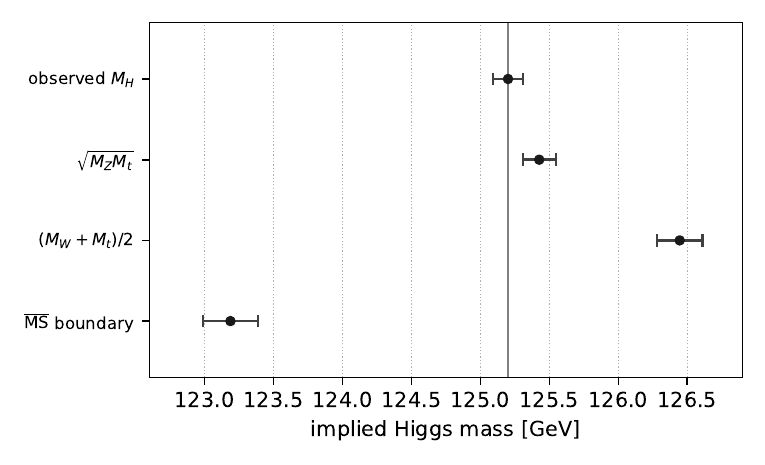}
\caption{Observed and implied Higgs masses.  The geometric pole relation is close to the observed value; the arithmetic relation is high; the exact $\overline{\rm MS}$ boundary at $\mu=M_t$ is low.}
\label{fig:masspred}
\end{figure}

\section{NNLO $\overline{\rm MS}$ matching}
\label{sec:matching}

At tree level,
\begin{equation}
  M_Z=\frac{g_Zv}{2},\qquad
  M_t=\frac{y_tv}{\sqrt2},\qquad
  M_H^2=2\lambda v^2 .
\end{equation}

Therefore the geometric mass relation corresponds, in a tree-level on-shell translation, to
\begin{equation}
  \lambda=\frac{g_Zy_t}{4\sqrt2}. 
  \label{eq:boundary}
\end{equation}

This equation is not a scheme-independent identity.  It becomes a precision statement only after matching pole observables onto running parameters.  We evaluate the complete NNLO weak-scale matching formulae of Refs.~\cite{Degrassi2012,Buttazzo2013}, using $\mu=M_t$ and the inputs of Table~\ref{tab:inputs}.  In the hypercharge convention $g_Y$ used in the SM covariant derivative, the numerical formulae are
\begin{align}
\lambda(M_t)&=0.12604+0.00206(M_H-125.15)
             -0.00004(M_t-173.34),\label{eq:match_lambda}\\
y_t(M_t)&=0.93690+0.00556(M_t-173.34)
          -0.00042\frac{\alpha_s(M_Z)-0.1184}{0.0007},\label{eq:match_yt}\\
g_2(M_t)&=0.64779+0.00004(M_t-173.34)
          +0.00011\frac{M_W-80.384}{0.014},\label{eq:match_g2}\\
g_Y(M_t)&=0.35830+0.00011(M_t-173.34)
          -0.00020\frac{M_W-80.384}{0.014},\label{eq:match_gy}\\
g_3(M_t)&=1.1666+0.00314\frac{\alpha_s(M_Z)-0.1184}{0.0007}
          -0.00046(M_t-173.34),\label{eq:match_g3}
\end{align}
where masses are in GeV.  We include the residual theoretical uncertainties quoted for $\lambda$ and $y_t$, $0.00030$ and $0.00050$, in the propagated errors.  This gives
\begin{align}
\lambda(M_t)&=0.12618\pm0.00038, & y_t(M_t)&=0.93258\pm0.00198,\nonumber\\
g_2(M_t)&=0.64764\pm0.00011, & g_Y(M_t)&=0.35842\pm0.00019,\nonumber\\
g_3(M_t)&=1.16518\pm0.00404.\label{eq:matched_values}
\end{align}
The matched version of Eq.~(\ref{eq:boundary}) is then tested by
\begin{equation}
  \rhomat(M_t)=\frac{\sqrt{g_2^2(M_t)+g_Y^2(M_t)}\,y_t(M_t)}{4\sqrt2\lambda(M_t)}.
  \label{eq:rho_ms_def}
\end{equation}
Numerically,
\begin{equation}
  \rhomat(M_t)=0.96714\pm0.00361,
  \qquad
  \ln\rhomat=-0.03342\pm0.00374.
  \label{eq:rho_ms_value}
\end{equation}

Thus the exact running-coupling boundary (\ref{eq:boundary}) at the top scale is not supported.  If Eq.~(\ref{eq:boundary}) is imposed while holding $M_t$, $M_W$ and $\alpha_s$ fixed, the matching equations predict
\begin{equation}
  M_H^{\msbar\,\mathrm{pred}}=123.19\pm0.20\GeV,
  \label{eq:mh_ms_pred}
\end{equation}
about $2\GeV$ below the observed Higgs mass.  Conversely, holding $M_H$, $M_W$ and $\alpha_s$ fixed gives
\begin{equation}
  M_t^{\msbar\,\mathrm{pred}}=177.81\pm0.50\GeV,
  \label{eq:mt_ms_pred}
\end{equation}
well above the direct top-mass average.  

The result is robust: the discrepancy is driven by the large finite threshold between pole masses and running couplings, especially in the top Yukawa coupling.

The finite threshold factor required to convert the matched running-coupling ratio into an exact low-energy boundary is
\begin{equation}
  \kappa_{\rm th}\equiv \rhomat^{-1}=1.0340\pm0.0039.
  \label{eq:kappa_target}
\end{equation}

If the aim is to reproduce the observed pole ratio rather than exact unity, the corresponding factor is $\rho_{Zt}/\widehat\rho_{Zt}=1.0377\pm0.0040$.  These numbers are useful because they turn the old qualitative question, ``is there a symmetry?'', into a quantitative matching target.

\begin{figure}[t]
\centering
\includegraphics[width=\columnwidth]{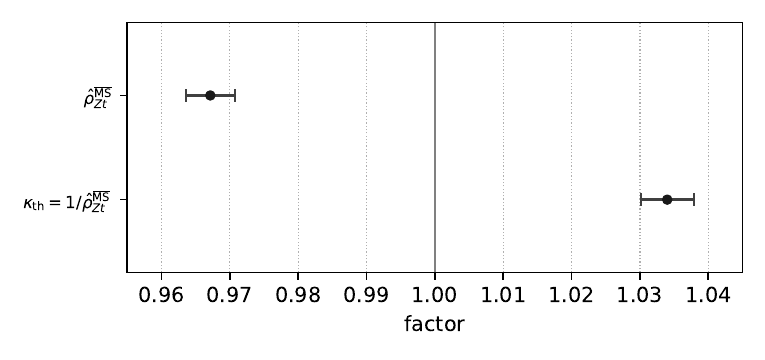}
\caption{The exact $\overline{\rm MS}$ boundary at $\mu=M_t$ fails by about $3.4\%$.  A symmetry acting at the running-coupling level must therefore supply, or predict, a finite threshold factor close to $\kappa_{\rm th}=1.034$.}
\label{fig:threshold}
\end{figure}

\section{Symmetry targets}
\label{sec:symmetry}

The matching result constrains possible interpretations.  A literal unbroken SM symmetry cannot enforce Eq.~(\ref{eq:boundary}), because $\lambda$, $y_t$ and $g_Z$ are independent renormalized couplings and their beta functions do not preserve $g_Zy_t/\lambda$ as an invariant.  The one-loop equations already show this: $y_t$ is strongly affected by QCD, whereas $\lambda$ is driven by a competition between $y_t^4$ and electroweak terms.  This is consistent with the standard near-criticality picture of the Higgs potential \cite{Degrassi2012,Buttazzo2013}.

A viable explanation must therefore be more structured.  Three possibilities are suggested by the numbers.

First, one may take the pole-level equation (\ref{eq:pole_relation}) as the primary relation.  Then the symmetry would not be a symmetry of the $\overline{\rm MS}$ Lagrangian parameters themselves, but a statement about physical thresholds.  In this picture the $3$--$4\%$ finite factor in Eq.~(\ref{eq:kappa_target}) is not a problem; it is part of the symmetry-breaking map from running couplings to pole observables.  The falsifiable prediction is then $M_t=171.90\pm0.30\GeV$ in the pole convention.

Second, one may postulate a broken top-Higgs custodial extension.  Ordinary custodial $SU(2)$ explains why the electroweak vector masses are organized by $M_W/M_Z$ up to hypercharge breaking, but it does not relate the scalar quartic to the top Yukawa coupling \cite{Sikivie1980}.  A larger spurionic symmetry could assign $g_Z$, $y_t$ and $\lambda$ to a common top-Higgs sector and enforce
\begin{equation}
  4\sqrt2\lambda=\kappa_{\rm th}\,g_Zy_t,
  \label{eq:kappa_sym}
\end{equation}
with the finite threshold in Eq.~(\ref{eq:kappa_target}).  Equation~(\ref{eq:kappa_sym}) is model-independent: any proposed custodial/top-Higgs symmetry must predict either $\kappa_{\rm th}=1$ at the pole level or $\kappa_{\rm th}\simeq1.034$ at the $\overline{\rm MS}$ level.

Third, a triality-like nonlinear symmetry could relate the heaviest scalar, fermion and neutral vector sectors only after electroweak symmetry breaking.  Such a symmetry cannot commute with the unbroken Lorentz and gauge structure in a conventional linear way; it would have to be realized through composite, pseudo-Nambu-Goldstone or threshold degrees of freedom.  Little-Higgs and composite-Higgs constructions show how quartics can be induced by gauge and top interactions, but they do not generically produce Eq.~(\ref{eq:kappa_sym}) without additional dynamics \cite{ArkaniHamed2002,Schmaltz2005}.  The present calculation fixes the additional dynamics required: it must account for a few-percent matching factor while leaving the pole-level relation at the per-mille-to-percent level.

In an effective-field-theory notation this requirement can be written as
\begin{align}
M_Z^2&=\frac{g_Z^2v^2}{4}(1+\delta_Z), &
M_t&=\frac{y_tv}{\sqrt2}(1+\delta_t),\nonumber\\
M_H^2&=2\lambda v^2(1+\delta_H).
\end{align}
Then
\begin{equation}
\frac{M_ZM_t}{M_H^2}=\frac{g_Zy_t}{4\sqrt2\lambda}
\left(1+\frac{\delta_Z}{2}+\delta_t-\delta_H\right)+O(\delta^2).
\label{eq:eft_threshold}
\end{equation}

The observed pattern therefore selects the threshold combination
\begin{equation}
  \frac{\delta_Z}{2}+\delta_t-\delta_H\simeq 0.038,
\end{equation}
if the matched running-coupling ratio is used as the starting point.  This is a precise target for SMEFT threshold fits or explicit ultraviolet completions \cite{Grzadkowski2010,Jenkins2013}.

\section{Conclusions}
\label{sec:conclusions}

The old Higgs mass coincidence has split into three sharply different statements.

The pole-level geometric relation $M_H^2\simeq M_ZM_t$ remains numerically interesting.  It predicts $M_t=171.898\pm0.302\GeV$ from $M_H$ and $M_Z$, and its present difference from the direct ATLAS--CMS average is about $1.4\sigma$ before adding any top-mass scheme ambiguity.

The arithmetic relation $2M_H\simeq M_W+M_t$ should not be used as an exact symmetry condition.  With present data it is a percent-level mnemonic, not a valid mass sum rule.

The exact $\overline{\rm MS}$ boundary condition $\lambda=g_Zy_t/(4\sqrt2)$ at $\mu=M_t$ is excluded by the NNLO matching evaluation.  It predicts $M_H\simeq123.19\GeV$ or $M_t\simeq177.81\GeV$, depending on which mass is solved for.  Therefore a viable theoretical explanation of the pole-level relation must involve finite threshold physics, a broken custodial/top-Higgs structure, or a nonlinear triality-like symmetry.  The quantitative target for such a construction is the matching factor $\kappa_{\rm th}=1.0340\pm0.0039$.

\section*{Acknowledgements}

The work of ETL has been supported in part by the Ministerio de Educaci\'on y Ciencia, grants FIS2025-24924,
 Universidad de Murcia project E024-018 and Fundacion Seneca (21257/PI/24). This manuscript extends and updates the analysis of Ref.~\cite{Torrente2014}.

\end{document}